\documentclass[12pt]{article}
\usepackage{amsfonts}
\usepackage{graphicx}

\newcommand{\eq}{\begin{equation}}
\newcommand{\eqx}{\end{equation}}
\newcommand{\eqn}{\begin{eqnarray}}
\newcommand{\eqnx}{\end{eqnarray}}
\newcommand{\f}[2]{\frac{#1}{#2}}
\newcommand{\qqqq}{\quad\quad\quad\quad}

\newcommand{\nn}{{\cal N}}
\newcommand{\zt}{\tilde{z}}
\newcommand{\dl}{\delta}
\newcommand{\Dl}{\Delta}
\newcommand{\cor}[1]{\left\langle{#1}\right\rangle}
\newcommand{\rsq}{{\mathfrak R^2}}

\title{Asymptotic perfect fluid dynamics as a consequence of AdS/CFT}

\author{Romuald A. Janik $^{a}$, Robi Peschanski $^b$\thanks{
e-mail: {\tt ufrjanik@if.uj.edu.pl}, {\tt pesch@spht.saclay.cea.fr}}\\ \\
$^a$ \small Institute of Physics, Jagellonian University,\\
\small Reymonta 4, 30-059 Krakow, Poland.\\
$^b$ \small CEA/DSM/SPhT Saclay
Unit\'{e} de Recherche associ{\'e}e au CNRS\\
\small CEA-Saclay, F-91191 Gif/Yvette Cedex, France.}

\date{}

\begin{document}

\maketitle

\begin{abstract}
We study the dynamics of  strongly interacting gauge-theory matter
(modelling quark-gluon plasma) in a boost-invariant setting using the
AdS/CFT correspondence. Using 
Fefferman-Graham coordinates and with the help of holographic
renormalization,  
we show that perfect fluid hydrodynamics emerges at large times as the
unique nonsingular asymptotic solution of the nonlinear Einstein
equations in the bulk. The gravity 
dual can be interpreted as a black hole moving off in the fifth
dimension. Asymptotic solutions different from perfect fluid behaviour
can be ruled out by the appearance of curvature singularities in the
dual bulk
geometry. Subasymptotic deviations from perfect fluid behaviour remain
possible within the same framework.   
\end{abstract}

\section{Introduction}

From the first years of the running of heavy ion collisions at RHIC,
evidence have been found that various observables are in good agreement
with models based on hydrodynamics \cite{hydro} and with
quark-gluon plasma (QGP) in a strongly coupled regime \cite{strongQGP}. 
To a large extent it seems that the QGP behaves approximately as a
perfect fluid as was first considered in \cite{Bjorken}. 

It is a challenge in QCD to derive from first principles the
properties of the dynamics of 
a strongly interacting plasma formed in heavy-ion collisions and in
particular to understand why the perfect fluid hydrodynamic equations
appear to be relevant.

Even if the experimental situation is still developing and rather
complex, it is worth simplifying the problem in order to
be able to attack it with appropriate theoretical tools.
Recently the AdS/CFT correspondence \cite{adscft,adscftrev} emerged as  a new
approach to 
study strongly coupled gauge theories. This has been largely worked
out in the supersymmetric case and in particular for the conformal
case of $\nn=4$ super Yang-Mills theory (SYM). Interestingly enough, since
the QGP is a deconfined and strongly interacting phase of QCD we could expect
that results for the nonconfining $\nn=4$ theory may be relevant. We
will make this assumption in our work.

The AdS/CFT correspondence has already been advocated in theoretical
studies in the context of heavy-ion collisions
\cite{son,nastase,sinzahed,zahedsin}. 
Transport coefficients at
finite temperature have been calculated using the static black hole
dual geometry and some generalizations \cite{son}, thermalization has
been suggested to be described by a black hole formation process
\cite{nastase}, and proposals have been put forward 
for the  gravity dual description of various processes during
heavy-ion collisions \cite{zahedsin} e.g. cooling as black hole motion
in the 5th direction.  

In this paper we focus on the spacetime evolution of the gauge theory
(4D) energy-momentum tensor, and derive its asymptotic behaviour from
the solutions of the nonlinear Einstein equations of the gravity
dual. 

Imposing the absence of curvature singularities in the gravity dual,
we will show that, in the boost invariant setting (as in
\cite{Bjorken}), perfect fluid hydrodynamics emerges from the AdS/CFT
solution at large times. The corresponding asymptotic solution of the
Einstein equations is given by formula (\ref{e.flgeom}) of our paper. 

The plan of our paper is as follows. In section 2 we review the
Bjorken hydrodynamics on the gauge theory side. Then, in section 3, we
setup a general framework of deriving a gravity dual for a given
energy-momentum tensor on the boundary, based on the
holographic renormalization method. 
In section 4 we derive the
large proper-time behaviour of the boost-invariant gravity duals by solving
analytically the corresponding nonlinear Einstein equations in the bulk. 
In section 5 we arrive at the physical solution by requiring the absence
of curvature singularities. This constraint selects perfect fluid
hydrodynamics in the 4D gauge theory. We close the paper with
conclusions and outlook.

\section{Bjorken hydrodynamics}

As is well known, a model of the central rapidity region
of heavy-ion reactions based on
hydrodynamics was pioneered in 
\cite{Bjorken} and involved the assumption of boost
invariance. 
In this paper we study the dynamics of strongly interacting
gauge-theory matter assuming boost
invariance.
Let us review now the picture which will serve as
a basis of our theoretical investigation.

We will be interested in the spacetime evolution of the
energy-momentum tensor $T_{\mu\nu}$ of the gauge-theory matter. It is
convenient to introduce proper-time ($\tau$) and rapidity ($y$)
coordinates in the longitudinal position plane:
\eq
x^0=\tau \cosh y \qqqq x^1=\tau \sinh y \ .
\eqx  
In these coordinates the Minkowski metric has the form
\eq
ds^2=-d\tau^2 +\tau^2 dy^2 +dx_\perp^2 \ .
\eqx
Assuming for simplicity, $y \to -y$ symmetry and
translational and rotational symmetry in the transverse plane, the
energy-momentum tensor has only three nonzero components
$T_{\tau\tau}$, $T_{yy}$ and $T_{x_2x_2}=T_{x_3x_3} \equiv T_{xx}$,
which depend
only on $\tau$. Since we are dealing with a {\em conformal} gauge
theory, $T_{\mu\nu}$ is necessarily traceless
\eq
\label{e.tr}
-T_{\tau\tau}+\f{1}{\tau^2} T_{yy}+2T_{xx}=0  \ .
\eqx
Energy-momentum conservation $D_\nu T^{\mu\nu}=0$ gives a
further relation between the components:
\eq
\label{e.cons}
\tau \f{d}{d\tau} T_{\tau\tau} +T_{\tau\tau}+\f{1}{\tau^2} T_{yy}=0
\eqx
So using relations (\ref{e.tr})-(\ref{e.cons}), all components of the
energy momentum tensor can be expressed in terms of a {\em single}
function $f(\tau)$: 
\eq
\label{e.tgen}
T_{\mu\nu}\! = \!
\left(\begin{tabular}{cccc}
$f(\tau)$ & 0 &0 & 0 \\
0 & $-\tau^3 \f{d}{d\tau} f(\tau)\!-\!\tau^2 f(\tau)$ & 0 & 0 \\
0 & 0 & $f(\tau)\!+\! \f{1}{2}\tau \f{d}{d\tau} f(\tau)$ & 0 \\
0 & 0 & 0 & $f(\tau)\!+\! \f{1}{2}\tau \f{d}{d\tau} f(\tau)$
\end{tabular}\right)
\eqx
where the matrix $T_{\mu\nu}$ is expressed in $(\tau,y,x_1,x_2)$
coordinates. 

Furthermore the function $f(\tau)$ is constrained by the positive
energy condition which states that for any {\em time-like} vector
$t^\mu$, the energy in the frame whose timelike axis is $t^\mu$ should
be positive i.e. 
\eq
T_{\mu\nu}t^\mu t^\nu \geq 0 \ .
\eqx
Using $t^\mu=(\sqrt{s^2+\tau^2 v^2+2w^2},v,w,w)$ we are led to the following  
restrictions 
\eq
\label{e.enpos}
f(\tau)\geq 0 \qqqq f'(\tau)\leq 0 \qqqq
\tau f'(\tau) \geq -4 f(\tau) \ .
\eqx

Note that all the above structure (\ref{e.tgen}) is purely based on
kinematics. The dynamics of the gauge
theory should pick a specific $f(\tau)$. A perfect fluid or a fluid with
nonzero viscosity and/or other transport coefficients will lead to
different choices of $f(\tau)$. 

The main aim of this paper is to address the problem of
determination of the function $f(\tau)$ from the AdS/CFT correspondence. 
Let us first describe two distinct cases of physical interest:

\subsubsection*{Perfect fluid: Bjorken hydrodynamics}

Let us assume that the gauge theory matter behaves like a perfect
fluid. This means that the energy-momentum tensor has the
form\footnote{Note that in ref.~\cite{Bjorken} the Minkowski metric has
$(+,-,-,-)$ signature instead of $(-,+,+,+)$ that we use.}
\eq
T_{\mu\nu}=(E+p) u_\mu u_\nu +p \eta_{\mu\nu}
\eqx 
where $u^\mu$ is the (local) 4-velocity of the fluid ($u^2=-1$), $E$
is the energy density and $p$ is the pressure. Boost invariant
kinematics then forces\footnote{In the $(\tau,y,x_1,x_2)$ coordinates.}
$u^\mu=(1,0,0,0)$ and the comparison with (\ref{e.tgen}) leads to
\eq
f(\tau)=\f{e}{\tau^{\f{4}{3}}}
\eqx
which is the result for the ideal relativistic fluid 
\cite{Bjorken} satisfying $E=3p$. Moreover the entropy per unit
rapidity remains constant, while the temperature cools down as 
\eq
T \sim \tau^{-1/3} \ .
\eqx

\subsubsection*{Free streaming case}

Let us now consider the {\it free streaming case}, where the longitudinal
pressure vanishes. This property is 
expected to be valid in the first stages of heavy-ion collisions when
the QCD coupling is small (see \cite{Kovchegov} for further comments).
Using (\ref{e.tgen}) this leads to
\eq
f(\tau)=\f{\tilde{e}}{\tau} 
\eqx 
where $\tilde{e}$ is a dimensionful constant.

In the following we will more generally introduce a family of $f(\tau)$
with the large $\tau$ behaviour of the form
\eq
f(\tau) \sim \tau^{-s} \ .
\eqx
Note that using the energy positivity constraint (\ref{e.enpos}) we
are led to consider\footnote{An interesting limiting case $s=4$
  satisfies also the constraint but requires a specific treatment
  which is beyond the scope of our paper.} $0<s<4$.

\section{Holographic renormalization}

Let us now turn to the AdS/CFT correspondence and describe the dual bulk
geometry corresponding to a given configuration of the gauge theory
energy-momentum tensor $T_{\mu\nu}$.

According to the AdS/CFT correspondence, vacuum expectation values of
(a class of) local operators in the gauge theory can be reconstructed from the
asymptotics of the dual supergravity fields near the boundary
\cite{KlebWit}. 
In the case of the energy-momentum tensor, the dual field is just the
metric. The 
reconstruction of the VEV $\cor{T_{\mu\nu}}$ from the near-boundary
asymptotics of the gravity solution has been first studied in
\cite{Balasubramanian,Myers}, and in a systematic way in ref.
\cite{Skenderis}.
 
Following \cite{Skenderis} we consider general asymptotically AdS
metrics in the so-called Fefferman-Graham coordinates \cite{Feff}:
\eq
\label{e.feff}
ds^2=\f{g_{\mu\nu} dx^\mu dx^\nu+dz^2}{z^2}
\eqx
Note that this choice leaves, in general, no remaining
diffeomorphism (coordinate) freedom since five conditions on the metric have
already been imposed. 

One then considers solutions of {\em vacuum} Einstein
equations\footnote{Note that there is no energy-momentum tensor in the
dual gravity construction.} with negative
cosmological constant $\Lambda=-6$ (which corresponds to standard
$AdS_5$ \cite{Skenderis}) and their expansion\footnote{In (\ref{e.gexp})
additional logarithmic terms might in principle appear (see \cite{Skenderis}),
yet we find them absent for the cases (Minkowski metric on the
boundary) considered in this paper.}
near the boundary at $z=0$: 
\eq
\label{e.gexp}
g_{\mu\nu}=g^{(0)}_{\mu\nu}+z^2 g^{(2)}_{\mu\nu}+z^4 g^{(4)}_{\mu\nu}+z^6
g^{(6)}_{\mu\nu}+\ldots \ .
\eqx
Here $g^{(0)}_{\mu\nu}$ is the {\em physical} 4D metric
for the gauge theory on the boundary. In the following we set it to the flat
Minkowski metric $g^{(0)}_{\mu\nu}=\eta_{\mu\nu}$. Then $g^{(2)}_{\mu\nu}$ is
found to be zero, while $g^{(4)}_{\mu\nu}$ is proportional to the VEV of the
energy momentum tensor\footnote{For a generic background metric on the
  boundary $g^{(0)}_{\mu\nu}$, the formula is
  more complicated~\cite{Skenderis}.}:
\eq
\label{e.giv}
\cor{T_{\mu\nu}}= const \cdot g^{(4)}_{\mu\nu} \ .
\eqx
Hence given some metric in the bulk, i.e. a solution of the
supergravity equations, the VEV of the gauge theory energy-momentum
tensor can be directly read off. 

In \cite{Skenderis} one also considers the inverse problem, namely how
to construct general supergravity solutions of the form
(\ref{e.feff}). It turns 
out that one has to give as inputs {\em both} $g^{(0)}_{\mu\nu}$ and
$g^{(4)}_{\mu\nu}$ in order to generate a solution. Einstein equations
impose on $g^{(4)}_{\mu\nu} \equiv \cor{T_{\mu\nu}}$  just two
consistency constraints : 
\eq
\cor{T^\mu_\mu}=0 \qqqq D_\nu \cor{T^{\mu\nu}}=0
\eqx
namely tracelessness (since we are in a conformal theory) and energy
momentum conservation\footnote{The covariant derivative here is the
  one for the gauge theory metric $g^{(0)}_{\mu\nu}$.} in the gauge
theory. Then from these data, the Einstein equations allow one to
recursively reconstruct in principle all higher 
terms $g^{(n)}_{\mu\nu}$ in (\ref{e.gexp}). This general procedure goes under
the name of {\it holographic renormalization} 
\cite{Skenderis,Skenderis2}.

It is crucial to note that Einstein equations by themselves do not impose {\em
any} further local constraints on $g^{(4)}_{\mu\nu}$, or equivalently on
$\cor{T_{\mu\nu}}$.
This raises a question about the predictive power of the
gauge/gravity correspondence in this context. It would seem that {\em
  a-priori} any conserved energy-momentum tensor gives a viable
gravity background. This would be unacceptable from the gauge theory
point of view, since the specific form of $\cor{T_{\mu\nu}}$ should be
determined by the gauge theory dynamics.

Therefore in order to proceed further we need to look for a global
condition which would allow us to determine a physically acceptable
solution and hence a physical energy-momentum
tensor profile selected among all formal possibilities.
One natural criterion is to require the absence of a `naked'
singularity in the bulk.  

In this respect we are inspired by the AdS/CFT correspondence for gauge
theories with $N_f \neq 0$ flavours \cite{nfrefs,mateos,Erdmanger}. There, the
embedding of a D7 brane is constructed with a coordinate $y_6$, 
behaving asymptotically\footnote{For precise definitions of the
variables and the geometrical setting see e.g. \cite{Erdmanger}.} for
$\rho \to \infty$ as   
\eq
\label{e.y6}
y_6=m+\f{c}{\rho^2}+\ldots
\eqx 
In Eq. (\ref{e.y6}) the leading term is the current quark mass $m$
which is fixed (alike  to the 
gauge theory metric $g^{(0)}_{\mu\nu}$ in our case), while the first
subleading term $c$ corresponds to the quark condensate
$\cor{\bar{\psi}\psi}$ (alike to the $g^{(4)}_{\mu\nu}$ term in
(\ref{e.gexp})). {\em A-priori} for fixed $m$ one can construct
locally an embedding for {\em any} condensate $c$. The requirement that
the embedding is {\em nonsingular} picks \cite{mateos,Erdmanger} the
unique physical value of the $\cor{\bar{\psi}\psi}$ condensate. 

We will see that similar reasoning can be applied for the case of
energy-momentum tensor. One can construct, order by order, a gravity
solution for any $f(\tau)$. The requirement of nonsingularity for the
dual geometry will allow to pick up the physical $f(\tau)$. 

Before we proceed to describe boost-invariant geometries let us
mention two examples where exact solutions of the Einstein equations
exist for certain {\em non boost-invariant} energy-momentum tensors.

\subsubsection*{Example I : the static black hole}

Let us consider a static isotropic energy momentum tensor with
$E=3p=const$. The corresponding geometry that we obtain from solving the
Einstein equations with the boundary condition (\ref{e.giv}) and a
metric of the Fefferman form
\eq
ds^2=\f{-A(z) dt^2 +B(z) dx^2}{z^2}+ \f{dz^2}{z^2}
\eqx
is
\eq
\label{e.bhfef}
ds^2=-\f{(1-z^4/z_0^4)^2}{(1+z^4/z_0^4)z^2} dt^2
+(1+z^4/z_0^4)\f{dx^2}{z^2}+ \f{dz^2}{z^2}
\eqx 
and the VEV of the energy momentum tensor can be read off from the
expansion of the metric (\ref{e.gexp})
\eq
\cor{T_{\mu\nu}}\propto g^{(4)}_{\mu\nu} =
\left(\begin{tabular}{cccc}
$3/z_0^4$ & 0 &0 & 0 \\
0 & $1/z_0^4$ & 0 & 0 \\
0 & 0 & $1/z_0^4$ & 0 \\
0 & 0 & 0 &  $1/z_0^4$
\end{tabular}\right) \ .
\eqx
The geometry (\ref{e.bhfef}) does not look very familiar at first
sight. However by performing a change of coordinates
\eq
\zt=\f{z}{\sqrt{1+\f{z^4}{z_0^4}}}
\eqx
we can see that it is exactly the standard AdS
static black hole
\eq
ds^2=-\f{1-\zt^4/\zt_0^4}{\zt^2} dt^2
+\f{dx^2}{\zt^2}+\f{1}{1-\zt^4/\zt_0^4} \f{d\zt^2}{\zt^2}
\eqx
with $\zt_0=z_0/\sqrt{2}$. In this way, via the Fefferman-Graham
coordinates we recover the result of \cite{Balasubramanian,Myers}.

For later reference let us quote the
Hawking temperature (equal to the gauge theory temperature)
\eq
T=\f{1}{\pi \zt_0} =\f{\sqrt{2}}{\pi z_0}
\eqx
and the entropy
\eq
S=\f{Area}{4 G_N^{(5)}} =\f{\zt^{-3}_0 V_3}{4 \cdot \f{\pi}{2}
  N^{-2}} = \f{\pi^2}{2} N^2 V_3 T^3   \ .
\eqx

\subsubsection*{Example II : the planar shockwave}

The second example of an exact solution using our method is the
geometry dual to a gauge theory shock-wave (on the boundary).
Shockwave solutions have been constructed in AdS
spaces \cite{ItzHor,other} whose sources were in
the bulk, or particles on the boundary \cite{Emparan}.

If we introduce light-cone coordinates $x^-=t-y$ and $x^+=t+y$ then the
background dual to the VEV of the energy-momentum tensor 
\eq
T_{--}=\mu\, \dl(x^-) 
\eqx 
is (in the Fefferman-Graham coordinates)
\eq
\label{e.shock}
ds^2=\f{-dx^- dx^+ + \mu z^4 \dl(x^-) d{x^-}^2 +dx^2_\perp}{z^2}+
\f{dz^2}{z^2} \ .
\eqx
One can check that this is an exact solution of the Einstein
equations\footnote{The same holds for the more general case where
  $\dl(x^-)$ is replaced by any function of~$x^-$.}. 

The metric (\ref{e.shock}) represents
the gravity background dual to a plane shell of matter moving at a speed of
light which is an interesting model of e.g. an
ultrarelativistically boosted large nuclei. This dual background
may be a good starting point to extend the study of saturation effects
to strong coupling (as it mimicks closely the setup at the origin of
the Colour Glass Condensate/JIMWLK picture \cite{BJIMWLK}). 

Ultimately one is interested in the collisions of two such
shockwaves approaching along two light-cone directions which is the
setting corresponding to heavy-ion reactions. 
We leave this difficult but interesting problem for
subsequent work. We will concentrate here on an idealized
boost-invariant description which would correspond to the description
of the central rapidity region \cite{Bjorken}.  

\section{Boost-invariant geometries}

Let us now come to the main issue of this paper namely the study of
dual geometries in the boost-invariant case.
We impose boost invariance, together with $y\to -y$ symmetry plus
translation and rotation invariance in the transverse plane. 
The most general form of the bulk metric respecting these symmetries in the
Fefferman-Graham coordinates reads 
\eq
\label{e.ansatz}
ds^2=\f{-e^{a(\tau,z)} d\tau^2 +\tau^2 e^{b(\tau,z)} dy^2
  +e^{c(\tau,z)} dx^2_\perp}{z^2} +\f{dz^2}{z^2}  \ .
\eqx
The three coefficient functions $a(\tau,z)$, $b(\tau,z)$ and
$c(\tau,z)$ must start off at small $z$ as $z^4$ according to
  (\ref{e.gexp})-(\ref{e.giv}) and (\ref{e.tgen}). In this paper we will
restrict ourselves to the energy density behaving like
\eq
\label{e.fpow}
f(\tau)=\f{1}{\tau^s}
\eqx
for $0<s<4$
and we concentrate on the resulting {\em leading} behaviour for $\tau
\to \infty$.  
Let us emphasize that
there is a lot of physical content also in the subleading behaviour
and this problem certainly deserves further study.  

First we solve the Einstein equations
\eq
\label{e.einst}
R_{\mu\nu}-\f{1}{2}g_{\mu\nu} R - 6\, g_{\mu\nu}=0
\eqx
order by order in $z$ as in (\ref{e.gexp}), starting from
(\ref{e.fpow}) and following the holographic
renormalization procedure. We have implemented the
iterative procedure using Maple \cite{Maple} to obtain exact
coefficients of the power series expansions like
\eq
\label{e.aexp}
a(\tau,z)=\sum_{n=0}^N a_n(\tau) z^{4+2n}
\eqx
to some order $N$. This method calls for comments.

On the one hand this form is difficult to use in
order to analyze possible singularities in the bulk since these occur
at the edge of the radius of convergence and it is difficult to
disentangle unambigously whether the effect comes from a finite radius
of convergence or is a mark of a genuine curvature singularity.

On the other hand, the knowledge of the power series solution helps
us to find the large $\tau$ asymptotics of the exact
solutions in an analytic form. Namely, by analyzing the structure of the power
series solutions (\ref{e.aexp}), we find that after introducing the
scaling variable 
\eq
v=\f{z}{\tau^{s/4}} 
\eqx 
the exact solutions behave like
\eqn
a(\tau,z) &=& a(v)+ {\cal O}\left(\f{1}{\tau^\#} \right) \nonumber \\
b(\tau,z) &=& b(v)+ {\cal O}\left(\f{1}{\tau^\#} \right) \nonumber \\
c(\tau,z) &=& c(v)+ {\cal O}\left(\f{1}{\tau^\#} \right) \ ,
\eqnx
where we denoted by `$\#$' a {\em positive} (here unspecified) power. 

In order to find $a(v)$, $b(v)$ and $c(v)$ in an analytical form we
insert the metric (\ref{e.ansatz}) into the Einstein equations
(\ref{e.einst}) and take the limit $\tau\to \infty$ keeping $v$
fixed. We obtain the following set of coupled nonlinear
equations:
\[
v(2 a'(v)c'(v)+a'(v)b'(v)+2 b'(v) c'(v)) -6a'(v)-6b'(v)
-12 c'(v)+ v c'(v)^2=0
\]
\[
 3v c'(v)^2+v b'(v)^2+2v b''(v)+4v c''(v)-6b'(v)-12 c'(v)+2v b'(v)
c'(v) =0 
\]
\[
2 v s b''(v) +2 s b'(v)+8 a'(v)-v s a'(v) b'(v) -8b'(v) +v s
b'(v)^2 + 
\]
\eq
4 v s c''(v) +4 s c'(v) -2 v s a'(v) c'(v) +2 v s c'(v)^2=0 \ .
\eqx
Taking a suitable linear combination of these equations and
integrating, we find that
the functions $a(v)$, $b(v)$, and $c(v)$ satisfy a linear relation
\eq
(4-3s) a(v)+(s-4)b(v)+2s c(v)=0 \ .
\eqx
After nontrivial transformations, the remaining equations may be solved
giving the solution 
\eqn
a(v) &=& A(v)-2m(v) \nonumber \\
b(v) &=& A(v)+(2s-2) m(v) \nonumber \\
\label{e.c}
c(v) &=& A(v)+(2-s) m(v)
\eqnx
where
\eqn
A(v)&=&\f{1}{2} \left( \log(1+\Dl(s)\,v^4) +\log(1-\Dl(s)\, v^4) \right) \\
m(v)&=&\f{1}{4\Dl(s)} \left( \log(1+\Dl(s)\,v^4) -\log(1-\Dl(s)\, v^4)
\right)
\eqnx
with
\eq
\Dl(s)=\sqrt{\f{3s^2-8s+8}{24}} \ .
\eqx
As a cross-check of this solution we have verified that performing a
power series expansion of (\ref{e.c}) indeed coincides
with the scaling $\tau \to \infty$ limit of the exact power series
solutions. 

Let us first specialize to
the two cases singled out in section 2, especially since the perfect
fluid case will turn out to be the only one physically relevant.

\subsubsection*{Perfect fluid case}

The perfect fluid corresponds to $s=4/3$ in (\ref{e.fpow}). Plugging
in this value in the above 
equations leads to the following {\em asymptotic} geometry
\eqn
\label{e.flgeom}
ds^2=\f{1}{z^2} \left[- \f{\left( 1-\f{e_0}{3}
      \f{z^4}{\tau^{4/3}}\right)^2}{1+\f{e_0}{3}\f{z^4}{\tau^{4/3}}} d\tau^2+
\left( {\textstyle 1+\f{e_0}{3} \f{z^4}{\tau^{4/3}}}\right) (\tau^2
      dy^2 +dx^2_\perp) 
\right] + \f{dz^2}{z^2}
\eqnx
where we reinstated the dimensionful parameter $e_0$ so that
$f(\tau)=e_0/\tau^{\f{4}{3}}$. 

Remarkably enough this geometry is of a form similar to the black hole
solution (\ref{e.bhfef}) but with the location of the horizon {\em moving}
in the bulk according to 
\eq
\label{e.hor}
z_0=\left(\f{3}{e_0} \right)^{\f{1}{4}} \cdot \tau^{\f{1}{3}} \ .
\eqx
From the similarity of the
geometry (\ref{e.flgeom}) to the black hole solution (\ref{e.bhfef}),
we may qualitatively infer the scaling of the temperature i.e.
\eq
T(\tau) \sim  \f{1}{z_0} \sim \tau^{-\f{1}{3}}
\eqx
and similarly for the entropy per unit rapidity and transverse area
\eq
S(\tau) \sim Area \sim \tau \cdot \f{1}{z_0^3} \sim const
\eqx
in agreement with Bjorken hydrodynamics.

A word of caution is necessary at this stage.
It is not clear whether it is possible in the general {\em evolving}
setting to identify in a precise way temperature and entropy of such
a geometry 
(see e.g. discussions in the context of dynamical horizons in general
relativity \cite{Ashtekar}). In addition, in the AdS/CFT context, a
change of coordinates in the 
bulk involving both $\tau$ and $z$ will modify which point of the
horizon lies `above' which point on the boundary thus making the
identification of a {\em local} temperature and entropy density
problematic. Nevertheless it is quite probable that some approximate
notions do exist. 

Finally let us note that our geometry (\ref{e.flgeom}) may be a
reliable tool to study gauge theory observables which are sensitive to
the bulk geometry not far away from the boundary.

In the next section we will indeed show that the solution
(\ref{e.flgeom}) is selected by a criterion of absence of curvature
singularities for large proper-times.

\subsubsection*{Free streaming case}

Inserting $s=1$ into the equations (\ref{e.c}) we find
that the resulting metric is no longer similar to a moving black hole even
with some different functional form of $z_0(\tau)$. Namely one gets
\eqn
ds^2&=&\f{ \Biggl( -(1+\f{v^4}{\sqrt{8}})^{\f{1-2\sqrt{2}}{2}}
  (1-\f{v^4}{\sqrt{8}})^{\f{1+2\sqrt{2}}{2}} dt^2+ 
 (1+\f{v^4}{\sqrt{8}})^{\f{1}{2}}
  (1-\f{v^4}{\sqrt{8}})^{\f{1}{2}} \tau^2 dy^2 +}{z^2} \nonumber\\ 
&&\f{+(1+\f{v^4}{\sqrt{8}})^{\f{1+\sqrt{2}}{2}}
  (1-\f{v^4}{\sqrt{8}})^{\f{1-\sqrt{2}}{2}} 
  dx^2_\perp \Biggr)}{z^2} +\f{dz^2}{z^2} 
\eqnx
where $v=z/\sqrt[4]{\tau}$.
It is qualitatively different from the perfect fluid case, in
particular it displays singularities or zeroes at
$v^4=\sqrt{8}$ in all coefficients. On a more
quantitative ground we will now perform an analysis of the
curvature properties of the whole above class of metrics for $0<s<4$.

\section{Singularities and curvature}

Looking at the general form of (\ref{e.c}) we see that
there is a potential singularity for $v=\Delta(s)^{-1/4}$. However as
is often the case in general relativity such a singularity may be
a purely coordinate singularity as indeed happens in the vicinity of
the static
black hole horizon. In order to unambiguously locate a physical singularity we
calculate a {\em scalar} invariant formed out of the Riemann
curvature tensor. The simplest one, the Ricci scalar
$R=g^{\mu\nu}R_{\mu\nu}$ is actually by definition equal to $-20$ for
{\em any} solution of the Einstein equations (\ref{e.einst}) as can be
directly calculated. 
Let us then calculate the square of the Riemann tensor
\eq
\rsq=R^{\mu\nu\alpha\beta}R_{\mu\nu\alpha\beta}
\eqx
as a probe of curvature singularities.

It turns out that we cannot directly reach
$v=\Delta(s)^{-1/4}$ for fixed $\tau$ since we have only an asymptotic
solution. Therefore we perform the calculations of $R$ and $\rsq$
by taking $\tau \to \infty$ while keeping $v$ fixed.

Performing first the calculations for $R$ we find that indeed
\eq
R=-20 + {\cal O}\left( \f{1}{\tau^\#} \right) \ .
\eqx
This shows that we can trust the leading asymptotic term and thus that
the asymptotic approximation is self-consistent.

\begin{figure}[t]
\centerline{\includegraphics[height=7cm]{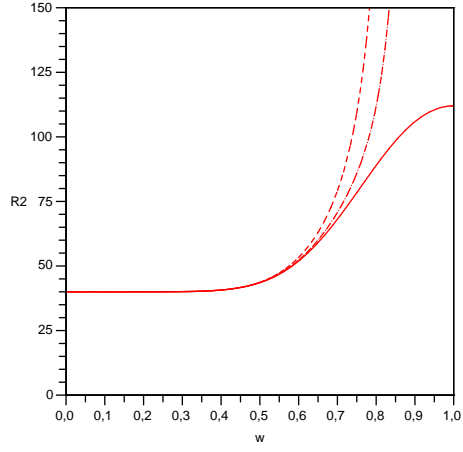}}
\caption{The curvature scalar $\rsq$ calculated as a function of
  $w=v/\Dl(s)^{1/4}$ for the perfect fluid case $s=4/3$ (solid line),
  $s=4/3-0.1$ (dotted line) and $s=4/3+0.2$ (dashed line).}
\end{figure}

The calculation of $\rsq$ gives a nontrivial result:
\eqn
\label{e.rsq}
\!\!\!\!&&\!\!\!\! \rsq =
\frac{4}{ \left( 1-
  { \Dl} \left( s \right)   ^{2}{v}^{8} \right) ^{4}}
\cdot \Biggl[ 10\,  { \Dl} \left( s \right)   ^{8}{v}^{32}
-88\,  { \Dl} \left( s \right)   ^{6}{v}^{24}+42\,{v}^{
24}{s}^{2}  { \Dl} \left( s \right)   ^{4}+ \nonumber\\
\!\!\!\!&& \!\!\!\! +112\,{v}^{24
}  { \Dl} \left( s \right)   ^{4}-112\,{v}^{24}  
{ \Dl} \left( s \right)   ^{4}s+36\,{v}^{20}{s}^{3}  {
 \Dl} \left( s \right)   ^{2}-72\,{v}^{20}{s}^{2}  {
 \Dl} \left( s \right)   ^{2}+ \nonumber\\
\!\!\!\!&& \!\!\!\! +828\,  { \Dl} \left( s
 \right)   ^{4}{v}^{16}+288\,{v}^{16}  { \Dl} \left( s
 \right)   ^{2}s-288\,{v}^{16}  { \Dl} \left( s
 \right)   ^{2}-108\,{v}^{16}{s}^{2}  { \Dl} \left( s
 \right)   ^{2}+ \nonumber \\
\!\!\!\!&& \!\!\!\! -136\,{v}^{16}{s}^{3}+27\,{v}^{16}{s}^{4}-320\,{
v}^{16}s+160\,{v}^{16}+296\,{v}^{16}{s}^{2}+36\,{v}^{12}{s}^{3} + 
\nonumber\\
\!\!\!\!&& \!\!\!\! -72\,{v}^{12}{s}^{2}
 -88\,  { \Dl} \left( s \right)   ^{2}{v}^{
8}+42\,{v}^{8}{s}^{2}+112\,{v}^{8}-112\,{v}^{8}s+10 \Biggr]
+ {\cal O}\!\left( \!\f{1}{\tau^\#}\! \right)
\eqnx
The striking fact is that in the range $0 <s<4$, the above
asymptotic expression for $\rsq$ diverges for $v=\Dl(s)^{1/4}$ for all
$s$ apart from the perfect fluid case $s=\f{4}{3}$ (see e.g.~fig.~1.). 
Remarkably enough, this result
requires a cancellation of the fourth order pole in front of
(\ref{e.rsq}). We checked this analytically by performing a Laurent
expansion near that pole. The final result for the perfect fluid case is
\eq
\rsq_{\mbox{\footnotesize{}perfect fluid}}=\f{8(5 w^{16}+20 w^{12}+174
w^8+20 w^4 +5)}{(1+w^4)^4} 
\eqx
where $w=v/\Dl(\f{4}{3})^{\f{1}{4}} \equiv \sqrt[4]{3}\, v$. $\rsq$ reaches
just a finite maximum of $\rsq=112$ at $v=1/\sqrt[4]{3}$, which plays
the role of the horizon, starting from the boundary value
of $\rsq=40$ (see fig. 1.).

This result means that for asymptotic times and for $s \neq \f{4}{3}$ one can
reach arbitrarily large curvatures in the bulk. This violates our
criterion of nonsingular bulk geometry. 
Thus we should conclude that the asymptotic behaviour for large $\tau$ of the
gauge theory energy-momentum tensor should be of the perfect fluid type.

Note that our result does {\em not} mean that we have an {\em exact} perfect
fluid. The above analysis was done only in the asymptotic $\tau \to
\infty$ regime. We showed that gauge theory dynamics rules out such
behaviours like streaming behaviour which have quite distinct
asymptotic $f(\tau) \sim 1/\tau^s$ with $s\neq \f{4}{3}$. It is quite
probable, however, that there would be subleading corrections due to
e.g. viscosity. In order to detect them one would have to
perform a more detailed analysis. This certainly
deserves further investigation.

\section{Conclusions and outlook}

Let us summarize our main results. 
\begin{itemize}
\item We propose a general framework for
studying the dynamics of matter (plasma) in strongly coupled gauge theory
using the AdS/CFT correspondence for the $\nn=4$ SYM theory.
\item We use tools related to holographic
renormalization for constructing dual geometries for given gauge theory
energy-momentum tensor profiles. We illustrate this method with the
static black hole and a planar shock wave solution.
\item Further imposing boost-invariant dynamics inspired by the Bjorken
  hydrodynamic picture, we derive the corresponding asymptotic
  solutions of the nonlinear Einstein equations. 
\item Among the family of asymptotic solutions, the only one with
  bounded curvature scalars is the gravity dual of a perfect fluid
  through its energy-momentum tensor profile.
\item This selected nonsingular solution, given by the metric
  (\ref{e.flgeom}), 
  is similar to a black hole moving off in the 5th dimension as a
  function of the physical proper time.   
\end{itemize}
Let us add some comments on the specific features of our approach and
results. In this paper we concentrate on looking for solutions of the
full nonlinear Einstein equations. It would be interesting to confront
this approach with the linearization methods of refs. \cite{son}. In
particular viscosity terms are expected to appear in the study of
subasymptotic terms. 
Note that the possibility of black hole formation in the {\em dual}
geometry has been argued in
ref. \cite{nastase}. 
More specifically, the geometry of a brane moving w.r.t. a black hole
background has been advocated in ref. \cite{zahedsin} for the dual
description of the cooling and expansion of a quark-gluon
plasma. In our case we could interpret the solution (\ref{e.flgeom})
as a kind of `mirror' situation in terms of a black hole moving off from
the AdS boundary. 
Note however that the precise geometrical identification of the
full solution would require further work, in particular for the
structure near the horizon and for subasymptotic proper times.   

The method and solution presented in this paper raise a lot of
stimulating questions for future investigation. 
In particular, one would like to address the key problem of connecting, on
general grounds, {\em local} physical temperature and
entropy of the gauge theory matter to features of the dual {\em evolving}
gravity solution. 

As a natural outlook it would
be interesting to study within the same framework possible deviations
from perfect fluid behaviour through subasymptotic gravity dual
solutions. It would be also interesting to somewhat relax
boost-invariance and study the possible corresponding modifications
involving rapidity dependence. In all cases we expect that
the condition of nonsingularity remains essential to select the proper
physical solution. Finally it would be stimulating to address the
initial problem of two colliding gauge-theory shock waves using
the framework of this paper.

\bigskip

\noindent{}{\bf Acknowledgments.} We thank Yuri Kovchegov for comments
on the manuscript.
RJ would like to thank the SPhT
Saclay for hospitality when this work was carried out. RJ was
supported in part by Polish Ministry of Science and Information
Society Technologies grants 2P03B08225 (2003-2006), 1P03B02427
(2004-2007) and 1P03B04029 (2005-2008).


\begin{thebibliography}{99}

\bibitem{hydro} P.~F.~Kolb and U.~W.~Heinz,
  ``Hydrodynamic description of ultrarelativistic heavy-ion collisions,''
  arXiv:nucl-th/0305084.

\bibitem{strongQGP} see e.g. E.~V.~Shuryak,
  ``What RHIC experiments and theory tell us about properties of  quark-gluon
  plasma?,''
  Nucl.\ Phys.\ A {\bf 750} (2005) 64
  [arXiv:hep-ph/0405066].

\bibitem{Bjorken}   J.~D.~Bjorken,
  ``Highly Relativistic Nucleus-Nucleus Collisions: The Central Rapidity
  Region,''
  Phys.\ Rev.\ D {\bf 27} (1983) 140.



\bibitem{adscft}   J.~M.~Maldacena,
  ``The large N limit of superconformal field theories and supergravity,''
  Adv.\ Theor.\ Math.\ Phys.\  {\bf 2}, 231 (1998)
  [Int.\ J.\ Theor.\ Phys.\  {\bf 38}, 1113 (1999)]
  [arXiv:hep-th/9711200];\\
  S.~S.~Gubser, I.~R.~Klebanov and A.~M.~Polyakov,
  ``Gauge theory correlators from non-critical string theory,''
  Phys.\ Lett.\ B {\bf 428}, 105 (1998)
  [arXiv:hep-th/9802109];\\
 E.~Witten,
  ``Anti-de Sitter space and holography,''
  Adv.\ Theor.\ Math.\ Phys.\  {\bf 2}, 253 (1998)
  [arXiv:hep-th/9802150].

\bibitem{adscftrev}
  O.~Aharony, S.~S.~Gubser, J.~M.~Maldacena, H.~Ooguri and Y.~Oz,
  ``Large N field theories, string theory and gravity,''
  Phys.\ Rept.\  {\bf 323}, 183 (2000)
  [arXiv:hep-th/9905111].



\bibitem{son}
  G.~Policastro, D.~T.~Son and A.~O.~Starinets,
  ``The shear viscosity of strongly coupled N = 4 supersymmetric Yang-Mills
  plasma,''
  Phys.\ Rev.\ Lett.\  {\bf 87}, 081601 (2001)
  [arXiv:hep-th/0104066];\\
  G.~Policastro, D.~T.~Son and A.~O.~Starinets,
  ``From AdS/CFT correspondence to hydrodynamics,''
  JHEP {\bf 0209}, 043 (2002)
  [arXiv:hep-th/0205052];\\
  P.~Kovtun, D.~T.~Son and A.~O.~Starinets,
  ``Viscosity in strongly interacting quantum field theories from black hole
  physics,''
  Phys.\ Rev.\ Lett.\  {\bf 94}, 111601 (2005)
  [arXiv:hep-th/0405231];\\
  A.~Buchel and J.~T.~Liu,
  ``Universality of the shear viscosity in supergravity,''
  Phys.\ Rev.\ Lett.\  {\bf 93}, 090602 (2004)
  [arXiv:hep-th/0311175].

\bibitem{nastase}
  H.~Nastase,
  ``The RHIC fireball as a dual black hole,''
  arXiv:hep-th/0501068.

\bibitem{sinzahed}
  S.~J.~Sin and I.~Zahed,
  ``Holography of radiation and jet quenching,''
  Phys.\ Lett.\ B {\bf 608}, 265 (2005)
  [arXiv:hep-th/0407215].

\bibitem{zahedsin}
  E.~Shuryak, S.~J.~Sin and I.~Zahed,
  ``A gravity dual of RHIC collisions,''
  arXiv:hep-th/0511199.


\bibitem{Kovchegov}   Y.~V.~Kovchegov,
  ``Isotropization and thermalization in heavy ion collisions,''
  arXiv:hep-ph/0510232.

\bibitem{KlebWit}   I.~R.~Klebanov and E.~Witten,
  ``AdS/CFT correspondence and symmetry breaking,''
  Nucl.\ Phys.\ B {\bf 556}, 89 (1999)
  [arXiv:hep-th/9905104].



\bibitem{Balasubramanian}   V.~Balasubramanian, J.~de Boer and D.~Minic,
  ``Mass, entropy and holography in asymptotically de Sitter spaces,''
  Phys.\ Rev.\ D {\bf 65}, 123508 (2002)
  [arXiv:hep-th/0110108].

\bibitem{Myers}  R.~C.~Myers,
  ``Stress tensors and Casimir energies in the AdS/CFT correspondence,''
  Phys.\ Rev.\ D {\bf 60}, 046002 (1999)
  [arXiv:hep-th/9903203].

\bibitem{Skenderis} S.~de Haro, S.~N.~Solodukhin and K.~Skenderis,
  ``Holographic reconstruction of spacetime and renormalization in the
  AdS/CFT correspondence,''
  Commun.\ Math.\ Phys.\  {\bf 217}, 595 (2001)
  [arXiv:hep-th/0002230].

\bibitem{Feff} C. Fefferman and C.R. Graham, ``Conformal Invariants,''
  in {\it Elie Cartan et les Math\'ematiques d'aujourd'hui},
  Ast\'erisque (1985) 95.

\bibitem{Skenderis2} K.~Skenderis,
  ``Lecture notes on holographic renormalization,''
  Class.\ Quant.\ Grav.\  {\bf 19}, 5849 (2002)
  [arXiv:hep-th/0209067].

\bibitem{nfrefs}  A.~Karch and E.~Katz,
  ``Adding flavor to AdS/CFT,''
  JHEP {\bf 0206}, 043 (2002)
  [arXiv:hep-th/0205236];

\bibitem{mateos}
  M.~Kruczenski, D.~Mateos, R.~C.~Myers and D.~J.~Winters,
  JHEP {\bf 0405}, 041 (2004)
  [arXiv:hep-th/0311270].

\bibitem{Erdmanger} J.~Babington, J.~Erdmenger, N.~J.~Evans,
  Z.~Guralnik and I.~Kirsch, 
  ``Chiral symmetry breaking and pions in non-supersymmetric gauge /  gravity
  duals,''
  Phys.\ Rev.\ D {\bf 69}, 066007 (2004)
  [arXiv:hep-th/0306018].


\bibitem{ItzHor} G.~T.~Horowitz and N.~Itzhaki,
  ``Black holes, shock waves, and causality in the AdS/CFT correspondence,''
  JHEP {\bf 9902}, 010 (1999)
  [arXiv:hep-th/9901012].

\bibitem{other}   G.~Arcioni, S.~de Haro and M.~O'Loughlin,
  ``Boundary description of Planckian scattering in curved spacetimes,''
  JHEP {\bf 0107}, 035 (2001)
  [arXiv:hep-th/0104039].

\bibitem{Emparan} R.~Emparan,
  ``Exact gravitational shockwaves and Planckian scattering on branes,''
  Phys.\ Rev.\ D {\bf 64}, 024025 (2001)
  [arXiv:hep-th/0104009].

\bibitem{BJIMWLK} For a recent review and references see
  E.~Iancu and R.~Venugopalan,
  ``The color glass condensate and high energy scattering in QCD,''
  arXiv:hep-ph/0303204.
 
\bibitem{Maple} Maple symbolic algebra program, Waterloo Maple Inc.

\bibitem{Ashtekar} A.~Ashtekar and B.~Krishnan,
``Isolated and dynamical horizons and their applications,''
Living Rev.\ Rel.\  {\bf 7}, 10 (2004)
[arXiv:gr-qc/0407042].



\end{thebibliography}
\end{document}